\documentclass[conference,compsoc]{IEEEtran}
\ifCLASSOPTIONcompsoc
  \usepackage[nocompress]{cite}
\else
  \usepackage{cite}
\fi
\ifCLASSINFOpdf
   \usepackage[pdftex]{graphicx}
\else
\fi
\usepackage{url}
\begin{document}
\title{A Dynamic Take on Window Management}
\author{\IEEEauthorblockN{Rohit Chouhan}
\IEEEauthorblockA{Application Security Engineer\\
PepsiCo, TX, US\\
\url{Email: rchouhan@cs.stonybrook.edu}}}
\maketitle
\begin{abstract}
On modern computers with graphical user interfaces, application windows are managed by a window manager, a core component of the desktop environment. Mainstream operating systems such as Microsoft Windows and Apple’s macOS employ window managers, where users rely on a mouse or trackpad to manually resize, reposition, and switch between overlapping windows. This approach can become inefficient, particularly on smaller screens such as laptops, where frequent window adjustments disrupt workflow and increase task completion time.

An alternative paradigm, dynamic window management, automatically arranges application windows into non-overlapping layouts. These systems reduce the need for manual manipulation by providing intelligent placement strategies and support for multiple workspaces. Despite their potential usability benefits, dynamic window managers remain niche, primarily available on Linux systems and rarely enabled by default.

This study evaluates the usability of dynamic window managers in comparison to conventional floating window systems.
We developed a prototype dynamic window manager that incorporates configurable layouts and workspace management, and we conducted both heuristic evaluation and statistical testing to
assess its effectiveness. Our findings indicate that dynamic window managers significantly improve task completion time in multi-window workflows by 37.83\%.

    By combining cognitive heuristics with empirical performance measures, this work highlights the potential of dynamic window management as a viable alternative to traditional floating window systems and contributes evidence-based insights to the broader field of human–computer interaction (HCI).
\end{abstract}
\IEEEpeerreviewmaketitle

\section{Introduction}
A window manager is a cornerstone of modern desktop environments. Since the advent of high-performance graphics and display technologies, windows have become the standard medium for two-way human–computer interaction (HCI) on personal computers. Early explorations of window managers, such as those described by Jensen \cite{jensen1992window}, established the fundamental principles that continue to shape contemporary desktop interfaces.

Conventional window managers, commonly found in systems such as Microsoft Windows, macOS, and mainstream Linux distributions, rely on overlapping or stacking windows that must be explicitly arranged by the user through point-and-click interaction. This model often leads to content being hidden beneath active windows, introducing potential inefficiencies in user workflows. Despite these limitations, the overlapping window paradigm has remained the dominant approach for decades.

In contrast, tiling window managers—where windows are automatically arranged to prevent overlap—have received comparatively little attention in mainstream computing. As Bly and Rosenberg \cite{bly} observed, the preference for overlapping windows is not grounded in evidence-based research. Their findings suggest that, for users with equivalent expertise, tiling window managers can outperform overlapping ones in both task completion time and the number of window operations required.

Building on this line of inquiry, the present study focuses on dynamic window managers. Unlike traditional tiling systems, dynamic window managers can automatically and intelligently adjust window placement based on configurable settings, offering users flexibility without requiring constant manual adjustments. These systems have gained popularity within the Linux community but remain under-studied in the broader HCI research landscape.

To address this gap, we evaluate dynamic window managers against conventional window managers through two complementary approaches: (1) a heuristic evaluation using Gerhardt-Powals’ cognitive engineering principles \cite{heuristic_evaluation}, and (2) statistical testing to measure performance differences in terms of task completion time. By combining cognitive and empirical perspectives, this research seeks to provide a more comprehensive understanding of the usability trade-offs between dynamic and conventional window management paradigms.

\section{Background}
\subsection{Traditional window managers}
Mainstream operating systems such as Microsoft Windows, Apple's macOS  ship with window managers that primarily employ overlapping, floating windows. While these systems provide limited support for tiling or snapping windows to specific regions of the desktop, they do not incorporate the full set of features found in dedicated tiling or dynamic window managers.
Conventional window managers are primarily mouse-driven, requiring frequent manual actions such as moving, resizing, and closing windows.

The efficiency of these actions can be understood through the lens of Fitts’s Law \cite{fitts}, which predicts that the time to acquire a target depends on the ratio between the distance to the target and the size of the target. In the context of window management, repeatedly pointing to and manipulating window borders or buttons introduces incremental delays that accumulate during workflows requiring frequent window operations. Over time, this contributes to measurable inefficiencies, particularly in scenarios that demand rapid interaction with multiple applications.

Although mainstream systems include limited snapping features (e.g., splitting windows into halves of the screen), the absence of robust shortcuts or more flexible layouts reduces their practical usability. As a result, desktops become crowded when many applications are open, and managing overlapping windows adds further cognitive and physical overhead, ultimately slowing down workflow performance.\\

\noindent
\includegraphics[width=\linewidth]{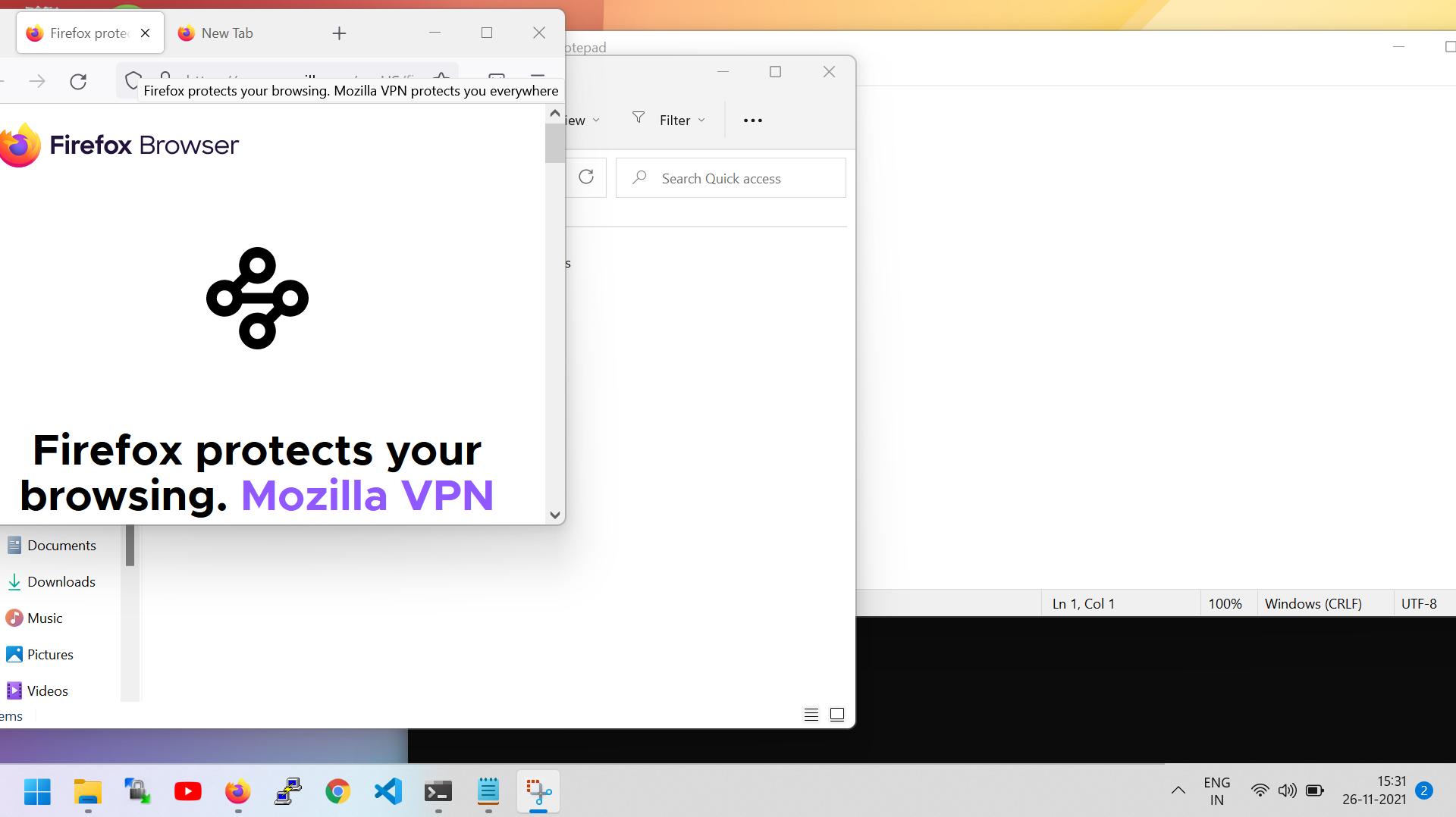}
{\small Figure 1 - Floating windows on Microsoft Windows}\\
\\
\noindent
\includegraphics[width=\linewidth]{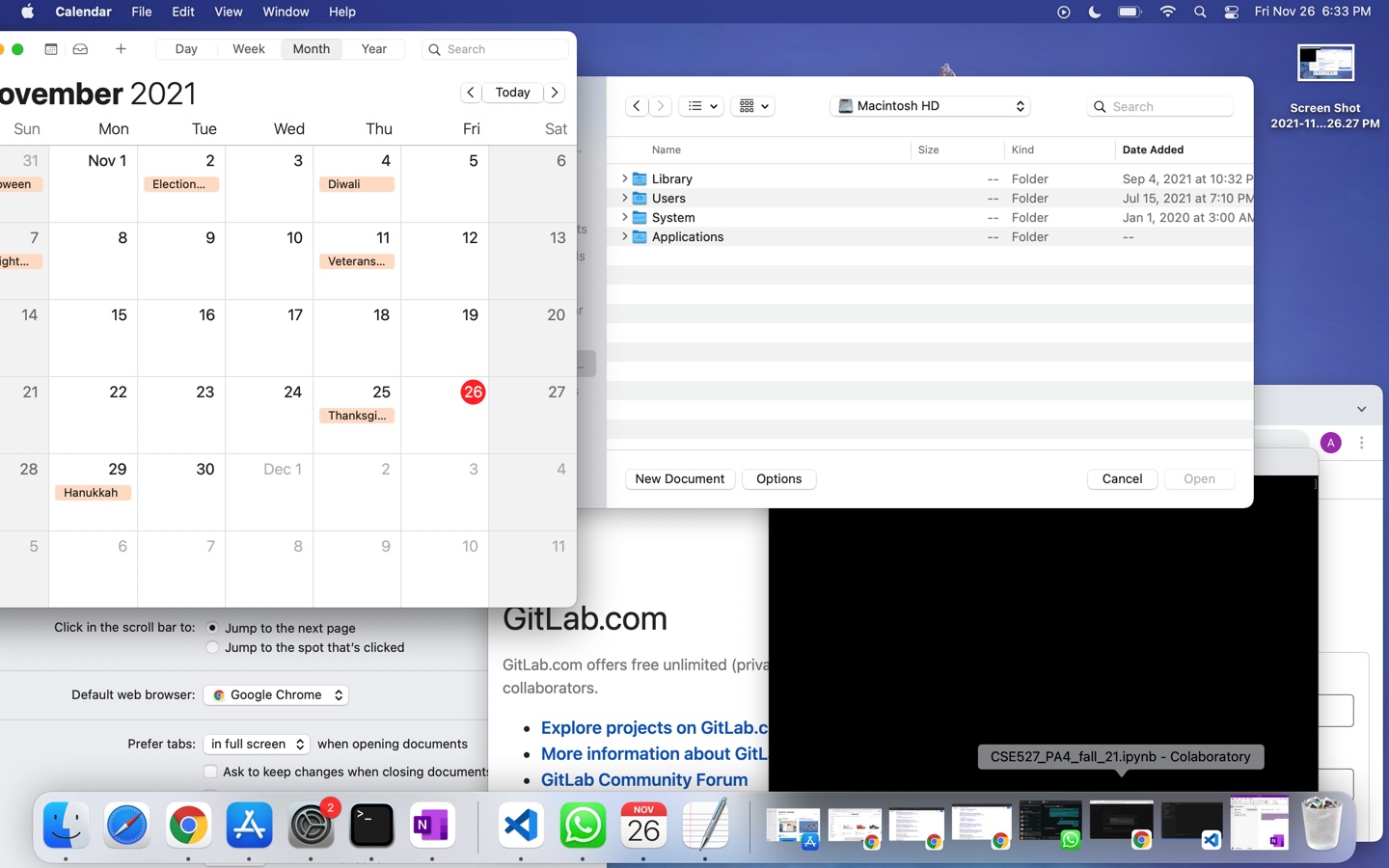}
{\small Figure 2 - Floating windows on macOS}

\subsection{Dynamic window managers}
Tiling window managers employ a fundamentally different model by automatically arranging windows into non-overlapping layouts. This approach reduces the need
for manual resizing and repositioning, offering users a more structured desktop environment. Despite evidence of their efficiency, tiling systems have remained underutilized, largely due
to their limited visibility outside Linux environments and their relatively steep learning curve. Deurzen \cite{vananatomy} provided a comprehensive study of tiling window managers in
the X Window System, detailing their implementation, operational modes, and the range of layouts available.
Such work highlights both the technical richness of tiling window managers and the barriers to their widespread adoption.

Building on these ideas, dynamic window managers represent a more recent development. These systems combine the benefits of tiling with enhanced configurability and intelligent placement
strategies. For example, Cohen \cite{cohen2007constraint} explored constraint-based tiling, where window placement, size, and relative positioning can be guided by formal rules,
offering a balance between automation and user control. Dynamic window managers extend this concept further by allowing users to define layouts, 
utilize multiple workspaces, and minimize reliance on constant manual adjustments.

The Linux ecosystem hosts a wide variety of dynamic window managers, including DWM, spectrWM, and Hyprland \cite{dynamic_window_manager}. While these systems have achieved
popularity among advanced Linux users and developers, they remain underexplored in the human–computer interaction research domain. In particular, few studies have
systematically compared dynamic window managers against conventional floating systems in terms of both cognitive heuristics and empirical performance outcomes.

\bigskip
\noindent
\includegraphics[width=\linewidth]{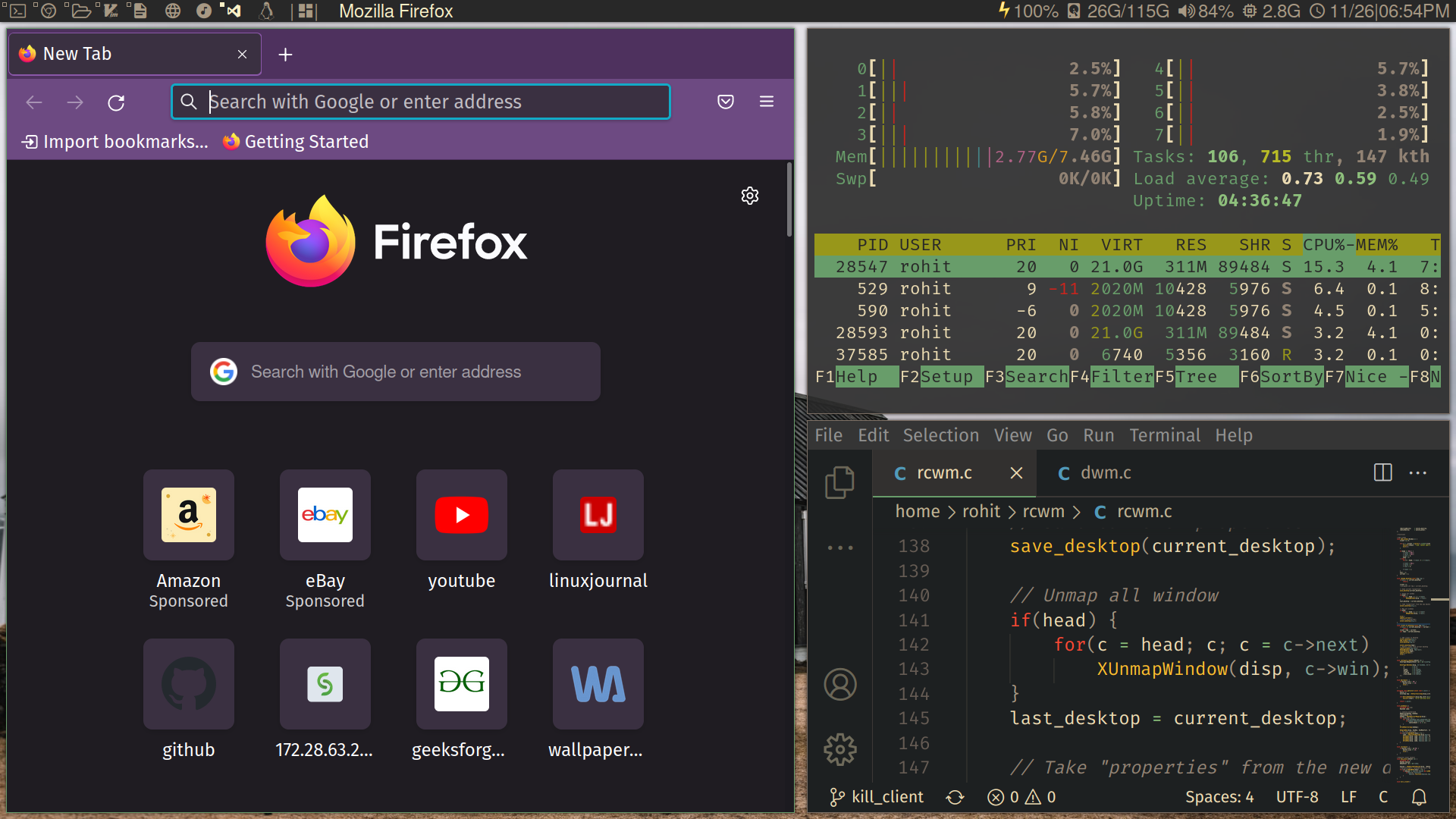}
{\small Figure 3 - DWM window manager on Linux}

\section{Design}
In this section we will be going over the design priciples and core components of a dynamic window manager. Through this we will understand what make a dynamic window management paradigm
different from traditional windows managemnent and how this different approach can bring benefits over traditional system.
\subsection{Workspaces}
Workspaces, sometimes referred to as virtual desktops, provide users with multiple logical screens to organize application windows. While the concept of workspaces
is not unique to dynamic window managers, these systems place greater emphasis on integrating them into everyday workflows. Rather than minimizing windows to a taskbar,
dynamic window managers encourage users to map applications to different workspaces, reducing visual clutter and simplifying task switching. Most dynamic window managers
also support integration with a status bar, which displays available workspaces and their active applications, thereby promoting spatial memory and reducing the
need for frequent window searches.

\subsection{Intelligent tiling and layouts}
Dynamic window managers automatically arrange application windows into layouts that maximize the use of available screen space. Among
the various layouts supported, the most common is the master–stack layout, illustrated in Figure 4. In this configuration, the most recently spawned window is
placed in the master area on the left side of the screen, while previously opened windows are stacked on the right in reverse chronological order. This approach ensures
that the primary task window remains prominent, while secondary windows remain accessible without requiring manual repositioning or resizing.
\begin{verbatim}
            --------------
            |        |___|
            |        |___|
            | Master |___|
            |        |___|
            |        |___|
            --------------
\end{verbatim}
{\small Figure 4 - Illustration of master and stack layout}

\subsection{Workspace and window rules}
Dynamic window managers provide mechanisms for automatically assigning specific applications to designated workspaces. In addition, individual application windows
can be launched with predefined states, such as floating mode, full-screen mode, or other user-specified configurations. Certain workspaces may also be bound to a
fixed layout or contain pinned windows that remain persistent across sessions. Collectively, these capabilities allow users to construct workflows that are not only highly
personalized but also consistent and efficient.

\subsection{Keyboard shortcuts}
Because dynamic window managers primarily adopt a keyboard-driven approach, nearly all window management actions can be mapped to customizable key combinations. These key
bindings are typically configured through a dedicated configuration file, though some implementations rely on external programs or utilities for managing them.

\section{Implementation}
To empirically compare conventional and dynamic window management paradigms, we developed a prototype dynamic window manager that implements the core functionality typical
of modern dynamic systems. The prototype was implemented in C, utilizing Xlib, the client-side library for the X11 windowing system \cite{xlib}.

Each workspace in the prototype maintains its own collection of application windows represented as a doubly linked list. When the user switches from one workspace to another,
the window manager first unmaps all currently visible windows, then traverses the linked list corresponding to the target workspace to map its windows onto the display.
This ensures efficient workspace isolation and fast transitions between contexts.

The tiling mechanism governs how windows are spatially organized on the screen. The prototype currently supports multiple layout modes, including master–stack and fullscreen.
In the master–stack layout (the default configuration), the most recently spawned window occupies a dominant region on the left side of the screen (the “master area”), while
previously opened windows are arranged in a vertical stack on the right. This approach maximizes visibility and accessibility of the active task while preserving
contextual awareness of secondary tasks.

When a new application is launched, the X11 window system notifies the manager of the new window creation event. The manager then invokes its tiling function to recompute and
update window geometries based on the active layout and the number of windows present. If only a single window exists within a workspace, it occupies the entire
screen area; additional windows trigger automatic rearrangement according to the current layout rules.

The system also provides keyboard-driven workspace management. Users can move the focused window to another workspace using predefined key bindings. When a
window is transferred, it is appended to the linked list associated with the target workspace. Switching between workspaces similarly relies on key bindings, allowing rapid,
mouse-free transitions.

Dynamic resizing is supported through an adjustable master area ratio. Users can modify the relative size of the master region using keyboard shortcuts, after which the tiling
function is re-executed to update window geometries. Each workspace maintains its own ratio, enabling flexible, context-dependent configurations.

Application termination is handled gracefully: when a user closes a window, the manager sends a quit signal to the corresponding client process and removes its entry
from the workspace’s linked list. Window mapping, unmapping, and destruction operations are all performed via native Xlib calls, ensuring compliance with X11’s event-driven windowing model.

The developed prototype serves as the experimental platform for this study. By implementing a simplified yet functional
dynamic window management environment, we enable a controlled comparison between conventional overlapping window systems and dynamic, keyboard-driven
interfaces. Participants will interact with both environments while completing predefined workflow tasks designed to simulate realistic multitasking scenarios.

\section{Evaluation}
In this section, we evaluate the usability and performance of dynamic window managers. The evaluation is conducted in two stages. First, we perform a heuristic evaluation based
on Gerhardt-Powals’ cognitive engineering principles \cite{heuristic_evaluation} to assess the interface design and identify usability strengths and weaknesses.
This qualitative analysis provides insights into how well dynamic window managers align with established human–computer interaction heuristics.

Following the heuristic evaluation, we conduct a quantitative analysis through controlled experiments to statistically compare the time taken to
complete equivalent workflows using both the conventional overlapping window manager and the dynamic window manager prototype.
The goal of this evaluation is to determine whether the dynamic window management approach offers measurable improvements in task efficiency and user experience.

\subsection{Hueristic evaluation}
A heuristic evaluation is a usability inspection method for computer software that helps to identify usability problems in the user interface (UI) design.
We will be using Gerhardt-Powals' cognitive engineering principles to assess the human-computer performance.

\noindent
The Gerhardt Powals' principles are listed below -
\begin{itemize}
    \item \textbf{Automate unwanted workflow} 
       
        Eliminate mental calculations, estimations, comparisons, and any unnecessary thinking, to free cognitive resources for high-level tasks.
    \item \textbf{Ruduce uncertainity}

        Display data in a manner that is clear and obvious to reduce decision time and error.
    \item \textbf{Fuse data}

        Bring together lower level data into a higher level summation to reduce cognitive load.
    \item \textbf{Present new information with meaningful aids to interpretation}

        New information should be presented within familiar frameworks
        (e.g., schemas, metaphors, everyday terms) so that information is easier to absorb.
    \item \textbf{Use names that are conceptually related to function}

        Display names and labels should be context-dependent, which will improve recall and recognition.
    \item \textbf{Group data in consistently meaningful ways}

        Within a screen, data should be logically grouped; across screens, it should be consistently grouped.
        This will decrease information search time.
    \item \textbf{Limit data-driven task}

        Use color and graphics, for example, to reduce the time spent assimilating raw data.
    \item \textbf{Include in the displays only that information needed by the user at a given time}

        Exclude extraneous information that is not relevant to current tasks so
        that the user can focus attention on critical data.
    \item \textbf{Provide multiple coding of data when appropriate}

        The system should provide data in varying formats and/or levels of detail in order to promote cognitive
        flexibility and satisfy user preferences.
    \item \textbf{Practice judicious redundancy}

        In order to be consistent, it is sometimes necessary to include more information than may be needed at a given time.
\end{itemize}

Evaluation of our dynamic window manager based on the principles listed above - 
\subsubsection{Automate unwanted workflow}
A dynamic window manager automates routine and repetitive tasks such as arranging windows on the screen, resizing, and repositioning them. Instead of manually dragging windows
with a mouse, users can rely on predefined keybindings to efficiently manage their workspace. Once the user becomes familiar with these keybindings, they no longer
need to search through a cluttered screen to locate a specific application. Instead, they can quickly switch to the appropriate workspace and resume their tasks seamlessly.

\subsubsection{Reduce uncertainity}
In a traditional window manager, new application windows typically spawn over existing ones or on the desktop, leading to visual overlap and reduced workspace
organization. In contrast, a dynamic window manager allows users to configure specific applications to open within designated workspaces. This capability reduces decision-making
time, minimizes user fatigue, and lowers the likelihood of errors by maintaining a structured and predictable workspace environment.

\subsubsection{Fuse data}
According to Gerhardt-Powals’ cognitive engineering principles, dynamic window managers embody the “fuse data” principle by integrating spatial, contextual, and
control information into coherent, unified representations. Through structured layouts, workspace-task mappings, and consolidated status indicators, these
systems reduce the need for users to mentally integrate disparate visual and contextual cues, thereby lowering cognitive load and improving situational awareness.

\subsubsection{Present new information with meaningful aids to interpretation}
Dynamic window managers effectively embody Gerhardt-Powals’ principle of presenting new information with meaningful aids to interpretation through their consistent, spatially
organized, and visually interpretable design. When a new application window is spawned, it is automatically placed according to a predictable layout, such as
the master-stack pattern, allowing users to immediately recognize its purpose and relationship to ongoing tasks. The spatial arrangement, window size, and status bar
indicators together provide contextual cues that reduce ambiguity and orient the user within their workspace. By ensuring that system changes are presented with clear visual
and contextual feedback, dynamic window managers help users interpret new information efficiently without breaking workflow continuity.

\subsubsection{Use names that are conceptually related to function}
In conventional window managers, system controls such as “maximize,” “close,” and “resize” are represented by labeled buttons or icons that are operated primarily
through mouse interaction. These visual controls use names and symbols that correspond to the action being performed, but they require additional pointer movement and
visual search. In contrast, dynamic window managers achieve conceptual clarity and functional consistency through configurable keybindings that directly map actions to
intuitive keyboard shortcuts. For instance, the maximize function can be mapped to Alt + M and toggling floating mode to Alt + F. This direct mapping of actions
to key combinations provides a clear and consistent relationship between the control and its function, reducing reliance on visual search and improving
interaction efficiency once the user learns the key mappings.

\subsubsection{Group data in consistently meaningful ways}
Dynamic window managers naturally embody the principle of grouping data in consistent and meaningful ways through the concept of workspaces. Each workspace acts as
a dedicated environment for related applications - for example, one workspace for coding, another for communication tools, and another for web browsing.
This organization allows users to mentally associate tasks with specific screen contexts, reducing cognitive load and search time. Unlike conventional window managers,
where multiple overlapping windows compete for visual attention, dynamic window managers maintain a structured layout where each window has a defined spatial position and purpose.
The consistent mapping of applications to specific workspaces enables users to quickly locate and switch to the relevant set of tools, fostering efficiency and minimizing distraction.

\subsubsection{Limit data-driven task}
Dynamic window managers align with the principle of limiting data-driven tasks by minimizing the user’s need to manually process and act upon visual data.
In conventional window managers, users must continuously scan the desktop, identify window positions, and decide where to move or resize them — all of which
are data-driven tasks requiring active interpretation. In contrast, dynamic window managers automate these spatial decisions through predefined layouts and intelligent window placement.
This abstraction allows users to focus on the task at hand rather than managing window geometry or screen real estate. By automating repetitive and visually intensive operations,
dynamic window managers convert low-level, data-oriented actions into higher-level interactions, thereby reducing cognitive fatigue and improving workflow efficiency.

\subsubsection{Include in the displays only that information needed by the user at a given time}
Dynamic window managers adhere to the principle of displaying only the information relevant to the user’s current task. By organizing
applications into distinct workspaces and using layouts that highlight the active window, the system ensures that users are presented primarily with
the tools they need at a given moment. Secondary windows are either hidden, stacked or placed in less prominent positions, reducing visual clutter and distraction.
Status bars and workspace indicators provide concise, contextually relevant feedback, such as which workspace is active or which window is focused,
without overwhelming the user with unnecessary details. This selective presentation of information allows users to concentrate on
the current task, improving efficiency and reducing cognitive load.

\subsubsection{Provide multiple coding of data when appropriate}
Dynamic window managers implement the principle of providing multiple coding of data by conveying system and window state through redundant and complementary cues. For example, the active window may be indicated not only by its position in the master area but also through visual borders, color highlighting, or status bar indicators. Workspaces are often represented simultaneously by numbers, labels, and visual icons in a status bar, allowing users to interpret workspace identity through multiple channels. This redundancy ensures that users can recognize relevant information quickly, even if one cue is missed or misinterpreted, thereby enhancing situational awareness and reducing the likelihood of errors.

\subsubsection{Practice judicious redundancy}
Dynamic window managers embody the principle of judicious redundancy by presenting critical information through multiple, carefully designed channels. For instance, the active window’s identity may be conveyed via its position in the master area, a highlighted border, and a status bar indicator showing the focused application. Similarly, workspaces are indicated by numbers, icons, and visual highlighting. By providing these redundant cues selectively and meaningfully, dynamic window managers ensure that users can quickly perceive essential information even if one channel is overlooked, while avoiding unnecessary visual clutter that could increase cognitive load

\subsection{Statistical testing}
To quantitatively assess the performance difference between the traditional and dynamic window managers, a within-subjects experimental design was used. Each participant completed an identical workflow task using both systems, and the time taken to complete the task was recorded for each condition.

Since the same participants were exposed to both conditions, a paired-samples t-test was performed to compare the mean task completion times between the two window managers. This test determines whether the observed difference in performance is statistically significant while accounting for individual variability.
\textbf{NULL Hypothesis}\\
The designed prototype is not faster as compared to a conventional window manager for achieving the same tasks on both systems.\\
\textbf{Alternate Hypothesis}\\
The designed prototype is faster as compared to a conventional window manager for achieving the same tasks on both systems and there is statistically significant difference
between mean of two groups.\\

\noindent
Independent variable - \textbf{Type of window manager}

\noindent
Dependent Variable - \textbf{Time taken to finish tasks}

\subsubsection{Experimental design}
A within-subjects experimental design was employed to evaluate the performance and usability differences between the traditional and dynamic window managers. Five participants were recruited for the study. Each participant was first briefed on the purpose of the experiment and the tasks they would be performing.

Participants were instructed to complete an identical set of workflow tasks using both systems: the traditional window manager (XFCE4), which follows a conventional overlapping window paradigm, and the developed dynamic window manager prototype. Prior to testing, participants were introduced to the
keybindings of the dynamic window manager and were given several practice attempts to familiarize themselves with the interface and keyboard-driven workflow.

The experiment was conducted on a Dell XPS 13 laptop running Linux to ensure consistent hardware and software environments across participants. The tasks were designed to simulate realistic, everyday use cases - such as opening multiple applications, organizing workspaces, and switching between tasks - to closely mirror real-world productivity scenarios.

Performance was measured by recording the time taken to complete each workflow under both window management systems. The collected data were analyzed using a within-subjects ANOVA \cite{anova}
test to determine whether there was a statistically significant difference in task completion times between the two systems.

\subsubsection{Experimental protocol}
The following experimental procedure was designed to simulate a realistic multitasking workflow commonly encountered by desktop users.
Each participant was asked to perform a sequence of actions on both the traditional window manager (XFCE4) and the dynamic window manager prototype, in the same order. The goal was to evaluate task efficiency and ease of window management.
Participant were asked to perform the following actions in order -
\begin{enumerate}
	\item Launch the Google Chrome browser.
	\item Open the file manager.
	\item Open a text editor.
	\item Arrange browser window and code editor window side by side.
	\item Navigate to x.org using the browser and copy the first line of text in the text editor.
	\item Quit both Chrome and text editor.
	\item Open two PDF documents side-by-side using a PDF viewer.
	\item Close the PDF viewer windows.
	\item Open a terminal, run \textit{ls} command, and then quit terminal.
\end{enumerate}

\section{Results}
The average task completion time across the five participants using the conventional window manager was 86.2 seconds, whereas
using the dynamic window manager prototype reduced the average time to 53.6 seconds. This represents an average time saving of approximately 32.6 (37.83\%
seconds per workflow, indicating a substantial improvement in efficiency.
Figure 5 illustrates the comparative results of the experiment, highlighting the performance advantage of the dynamic window manager over the conventional system.

\includegraphics[width=\linewidth]{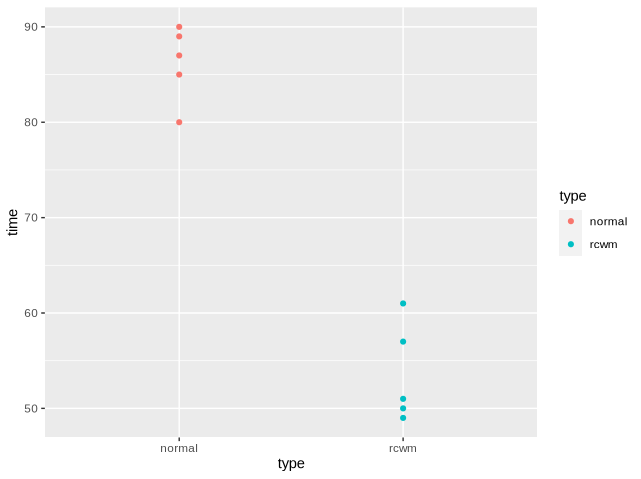}
{\small Figure 5 - Results of the experiment}\\

We followed the ANOVA test with a pairwise t-test comparing the two groups—normal (conventional window manager) and rcwm (our dynamic window manager prototype). The test was performed using a pooled standard deviation (SD), which assumes that both groups have equal variances and uses a combined estimate of variability. This method allows for a direct comparison of mean task completion times between the two systems to determine whether the observed performance difference is statistically significant.
\begin{footnotesize}
\begin{verbatim}
	Pairwise comparisons using t tests with pooled SD

data:  data$time and data$type

     normal
rcwm 3.7e-06

\end{verbatim}
\end{footnotesize}
As the P value of this test is 3.7e-06 which is much less than the threshold of 0.05, so we can reject the NULL hypothesis and accept the alternate hypothesis that
the time difference between the two different types of window managers is not due to variance.

\includegraphics[width=\linewidth]{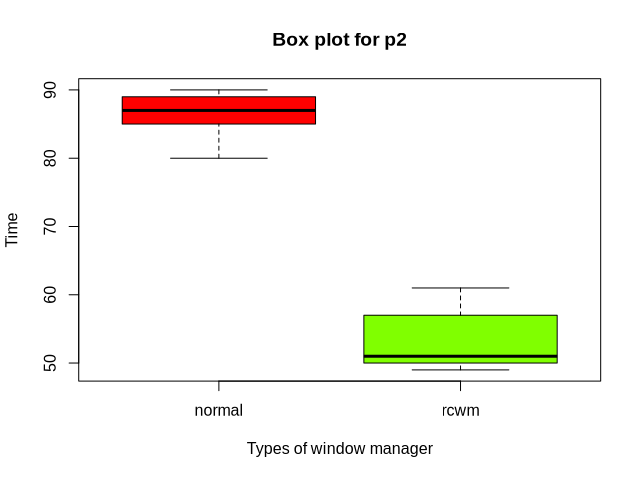}
{\small Figure 6 -A box plot of the results}

\section{Discussion}
The results of the statistical test indicate that the proposed dynamic window manager prototype produced a significant improvement in task completion time, reducing the average workflow duration by 32.6 seconds compared to the conventional window manager.
In addition to measurable performance gains, the prototype also demonstrated strong alignment with several of Gerhardt-Powals’ cognitive engineering heuristics, suggesting improvements in usability and workflow efficiency.

However, dynamic window managers are not without limitations. They tend to rely more heavily on recall rather than recognition, which can present a learning curve for new users. To promote broader adoption and enhance usability, we propose the following design improvements:

\begin{itemize}
    \item Introduce an on-screen widget or cheat sheet displaying commonly used keybindings. 
    \item Provide a quick-access overlay to reveal window control buttons when needed. 
    \item Include an interactive guided tour during first-time setup to familiarize users with controls and workflows. 
    \item Display contextual suggestions and tips on the desktop to remind users of relevant keybindings. 
    \item Offer a graphical settings panel to configure window manager preferences without requiring manual configuration file edits.

\end{itemize}

\section{Conclusion and future work}
Dynamic window manager show a quantifiable improvement in workflow speed over conventional window managers. After incorporating some of the recommended improvements they can
become very approachable for a wider audience. We strongly belive that their paradigm of managing windows will benefit a larger userbase, especially power users, programmers, or designers
who value a smoother and faster workflow. The current window management system popularized by macOS and Windows has been based on the same principles. There is a need to try a new
paradigm to make improvements in the human computer interaction.

\newpage
\bibliographystyle{IEEEtran}
\bibliography{references}
\end{document}